\documentclass{article}
\usepackage{spconf,amsmath,graphicx}
 \usepackage[T1]{fontenc}  % New fonts (preferred)
\usepackage{amsmath,amssymb}
\usepackage{graphicx}
\usepackage{pstricks}
\usepackage{psfrag}
\input alphabet
\def\diag{{\mathrm{diag}}}

\newsavebox{\fminibox}
\newlength{\fminilength}
\newenvironment{fminipage}[1][\linewidth]
  {\setlength{\fminilength}{#1}%-2\fboxsep-2\fboxrule}%
   \begin{lrbox}{\fminibox}\begin{minipage}{\fminilength}}
  {\end{minipage}\end{lrbox}\noindent\fbox{\usebox{\fminibox}}}

  \def\+{^\dagger}
\def\nequiv{\not\kern-.05em\equiv}
\def\egal{\kern-.5em=\kern-.5em}        % Moins d'espace autour de "="
\def\propt{\kern-.2em\propto\kern-.2em} % Idem
\def\intdouble{\int\kern-0.3em\int}
\def\inttriple{\int\kern-0.3em\int\kern-0.3em\int}

\def\rond#1{\overset{\kern-0.33em~_\circ}{#1}}
\def\rondit[#1]#2{\overset{\kern#1~_\circ}{#2}}

\def\babs{\begin{abstract}}             \def\eabs{\end{abstract}}
\def\barr{\begin{array}}                \def\earr{\end{array}}
\def\bcc{\begin{center}}                \def\ecc{\end{center}}

\def\bdes{\begin{description}}          \def\edes{\end{description}}
\def\bdoc{\begin{document}}             \def\edoc{\end{document}}
\def\ben{\begin{enumerate}}             \def\een{\end{enumerate}}
\def\beqn{\begin{eqnarray}}             \def\eeqn{\end{eqnarray}}
\def\beqnl#1{\beqn\label{#1}}           \def\eeqnl#1{\label{#1}\eeqn}
\def\beqnx{\begin{eqnarray*}}           \def\eeqnx{\end{eqnarray*}}
\def\bseqn{\begin{subeqnarray}}         \def\eseqn{\end{subeqnarray}}
\def\beq#1\eeq{\begin{equation}#1\end{equation}}
\def\bal#1\eal{\begin{align}#1\end{align}}
\def\balx#1\ealx{\begin{align*}#1\end{align*}}
\def\beqx{$$}                           \def\eeqx{$$}
\def\bfig{\protect\begin{figure}}       \def\efig{\protect\end{figure}}
\def\bfigx{\protect\begin{figure*}}     \def\efigx{\protect\end{figure*}}
\def\bfigt{\protect\begin{figurette}}   \def\efigt{\protect\end{figurette}}
\def\bfl{\begin{flushleft}}             \def\efl{\end{flushleft}}
\def\bfr{\begin{flushright}}            \def\efr{\end{flushright}}
\def\bit{\begin{itemize}}               \def\eit{\end{itemize}}
\def\bmi{\begin{minipage}}              \def\emi{\end{minipage}}
\def\bfmi{\begin{fminipage}}            \def\efmi{\end{fminipage}}
\def\bpic{\begin{picture}}              \def\epic{\end{picture}}
\def\bqu{\begin{quote}}                 \def\equ{\end{quote}}
\def\bqun{\begin{quotation}}            \def\equn{\end{quotation}}
\def\bsl{\begin{slide}}                 \def\esl{\end{slide}}
\def\btabb{\begin{tabbing}}             \def\etabb{\end{tabbing}}
\def\btabl{\begin{table}}               \def\etabl{\end{table}}
\def\btablx{\begin{table*}}             \def\etablx{\end{table*}}
\def\btab{\begin{tabular}} %\def\etab{\end{tabular}} Conflit avec eta gras... Elle est pas grasse, ma soeur !
\def\btabu{\begin{tabular}}             \def\etabu{\end{tabular}}
\def\btabx{\begin{tabular*}}            \def\etabx{\end{tabular*}}
\def\bbib{\begin{thebibliography}{999}} \def\ebib{\end{thebibliography}}
\def\bver{\begin{verbatim}}             \def\ever{\end{verbatim}}
\def\bca{\begin{cases}}                  \def\eca{\end{cases}}

\def\pardef{:=}%\stackrel{\textit{\em def}}{=}}
\def\ie{\textit{i.e.}\XS}
\def\eg{\textit{e.g.}\XS}
\def\prox#1{\Pc_{#1}}

\graphicspath{{fig1/}{fig2/}}

\usepackage{soul} 

\def\remSB#1{{\color{red}\footnote{{\color{red}(Remarque de SB~: #1)}}}}
\def\addSB#1{{\color{red} #1}}
\def\suppSB#1{{\color{red}\st{\footnotesize #1}}}

\def\addJI#1{{\color{blue} #1}}

\title{Fluorescence blind structured illumination microscopy: A new reconstruction strategy}

 \name{S. Labouesse$^{1}$, M. Allain$^{1,\star}$,  J. Idier$^{2}$,
   S. Bourguignon$^{2}$, A. Negash$^{1}$, P. Liu$^{2}$, A. Sentenac$^{1}$ }

\address{
  $^{1}$ Aix-Marseille Universit\'e, CNRS, Centrale Marseille, Institut Fresnel (France) \\
  $^{2}$ \'Ecole Centrale de Nantes, CNRS, IRCCyN (France)\\[1em]
  $^{\star}$ Corresponding author (marc.allain@fresnel.fr)
}

\begin{document}
\maketitle

\begin{abstract}
In this communication, a fast reconstruction algorithm is proposed for
fluorescence \textit{blind} structured illumination microscopy (SIM) under the sample positivity constraint. 
This new algorithm is by far simpler and faster than existing solutions, paving
the way to 3D and real-time 2D reconstruction.
\end{abstract}


\section{Introduction}

Classical wide-field fluorescence microscopy aims at imaging the 
fluorescence density $\rho$ emitted from a marked biological sample. 
In the linear regime, the recorded intensity is related to $\rho$ 
\textit{via} a simple convolution model~\cite{Goodman05}. If one proceeds to 
$M$ distinct acquisitions, the dataset $\{y_m\}_{m=1}^M$ is given by
\begin{equation}
  y_m  =  h  \otimes (\rho \times I_m) + \varepsilon_m \qquad m = 1\cdots M
  \label{eq:observation}
\end{equation}
where $\otimes$ is the convolution operator, $h$ is the point-spread 
function (PSF),  $I_m$ is the $m$-th illumination intensity pattern, and 
$\varepsilon_m$ is a perturbation term accounting for (electronic) noise
in the detection and model errors. The final resolution of the microscope is 
ultimately limited by the optical transfer function (OTF) whose cutoff frequency is fixed by the emitted wavelength 
and by the numerical aperture of the microscope objective. 
However, if any frequency component above this limit cannot be 
measured with uniform illuminations, structured illuminations 
can be used to shift high-frequency components of the object 
into the OTF support~\cite{Heintzmann99}. Such a strategy results in the standard 
\textit{structured illumination microscopy} (SIM) that resorts to 
harmonic illumination patterns to achieve super-resolution 
reconstruction. Because SIM uses the illumination patterns as 
references, strong artifacts are induced if the patterns are not
known with sufficient accuracy~\cite{Mudry12,Ayuk13}. From a practical viewpoint, 
such a condition is very stringent and restricts standard SIM to 
thin samples or to samples with small refraction indices~\cite{Mudry12}. 

The Blind-SIM strategy~\cite{Mudry12} has been proposed 
to tackle this problem, the principle being to retrieve the sample
fluorescence density without the knowledge of the 
illumination patterns, thereby extending the potential of SIM. 
In addition, this strategy promotes the use of speckle illumination 
patterns instead of harmonic illumination patterns, the latter
standard case being much more difficult to generate.   
From the methodological viewpoint, Blind-SIM relies on the \textit{simultaneous} 
reconstruction  of the fluorescence density and of the illumination patterns. 
More precisely, in~\cite{Mudry12}, joint reconstruction is achieved through the iterative resolution of
a constrained least-squares problem based on conjugate gradient 
iterations. However, the computational time of such a scheme (as 
reported in~\cite[Supplementary material]{Mudry12}) clearly restricts 
the applicability of the resulting joint blind-SIM strategy.
In this paper, the implementation issues of joint Blind-SIM are revisited 
and drastically simplified: a much improved implementation is
proposed, with an execution time decreased by several orders of 
magnitude. Moreover, it can be highly parallelized, opening the 
way to real-time reconstructions.

A few other contributions advocated the use of speckle patterns 
for super-resolved imaging in fluorescent \cite{Min13,Oh13} or
photo-acoustic \cite{Chaigne16} microscopy.   Because these strategies 
are derived from the statistical modeling of the speckle, the
resulting super-resolution strongly rests on the random nature 
of the illumination patterns \cite{Idier15a}. In contrast, this
communication stresses that the super-resolution mechanism 
behind joint blind-SIM relies on a sparsity and positivity
constraint enforced by the illumination pattern. As a result, 
super-resolved reconstructions can be obtained with a wide range 
of illuminations patterns (even with deterministic illuminations) as 
soon as they  cancel-out `` frequently''  the object.

\section{Blind-SIM problem reformulation}
\label{Reformulation}
In the sequel, we focus on a discretized formulation of the observation model 
\eqref{eq:observation}.
Solving the two-dimensional (2D) joint Blind-SIM reconstruction 
problem is equivalent to finding a \textit{joint} solution  $(\widehat\rhob,\{\widehat{\Ib}_m\}_{m=1}^M)$ to the
following constrained minimisation problem~\cite{Mudry12}:
\begin{subequations}
  \label{critere1}
  \begin{align}
  \label{critere1a}
   \min_{\rhob ,\{\Ib_m\}} \quad  {\textstyle \sum_{m=1}^{M}}\, \left\| \yb_m -  \Hb\diag(\rhob)\,\Ib_m\right\|^2&&\\[0em]
  \label{critere1b}
   \text{subject to}   \qquad {\textstyle \sum_m} \,\Ib_{m} \, =  \, M
   \times \Ib_{0} \qquad \qquad\\[0em]
   \label{critere1c}
    \text{and}\quad\,  \rho_n \geq 0, \quad I_{m;n} \geq 0, \qquad  \forall m,n 
%    \nonumber
  \end{align}
\end{subequations}
with $\Hb\in\eR^{N\times N}$ the 2D convolution matrix built from the
discretized PSF. We also denote $\rhob=\textbf{vect} (\rho_n)\in \eR^N$
the discretized fluorescence density,  $\yb_m=\textbf{vect} (y_{m;n})\in \eR^N$ 
the $m$-th recorded image, and  $\Ib_m=\textbf{vect}(I_{m;n})\in \eR^N$ the $m$-th
illumination with spatial mean $\Ib_0=\textbf{vect} (I_{0;n})\in \eR^N$. 
Note that \eqref{critere1} is a \textit{biquadratic} problem, which
was tackled in \cite{Mudry12} by a rather complex iterative scheme. 
However, problem \eqref{critere1} has a very specific structure that 
can benefit from a specific optimization strategy, which is explained hereafter.

% Let us remark that \eqref{critere1} is a \textit{biquadratic}
% problem. Block coordinate descent alternating between the object 
% and the  illuminations could be a possible minimisation strategy, 
% relying on sequentially solving $M+1$ quadratic programming problems.
% In~\cite{Mudry12}, a more efficient but more complex scheme is proposed. However,
% %belongs to the class of \textit{biquadratic} problems [REF] which are known to be non-convex in general.
% the minimisation problem \eqref{critere1} has a very specific
% structure, yielding a fast and simple strategy, as shown below.
%\bigskip

\subsection{A reformulation of the joint blind-SIM strategy}
Let us first consider problem \eqref{critere1} without the equality
constraint \eqref{critere1b}. It then becomes equivalent to $M$ quadratic minimisation problems 
\begin{subequations}
  \label{critere2}
  \begin{align}
    \label{critere2a}
    {\textstyle \min_{\qb_m} }%\quad {\textstyle \sum_{m=1}^{M}}\,
		\left\| \yb_m -  \Hb\qb_m\right\|^2 &&\\[0em]
    \label{critere2b}
    \text{subject to} \quad \qb_m\geq 0 \qquad \qquad \qquad 
  \end{align}
\end{subequations}
with $\qb_m \pardef \textbf{vect}(\rho_n\times I_{m;n})$.
Each minimisation problem \eqref{critere2} can 
be solved in a simple and efficient way 
(see Sec.~\ref{algo}), hence providing a set of global minimizers 
$\{\widehat{\qb}_m\}_{m=1}^M$. Although the latter set corresponds 
to an infinite number of solutions
$(\widehat\rhob,\{\widehat\Ib_m\}_{m=1}^M)$, 
the equality constraint in \eqref{critere1b} defines a unique solution such that 
$\diag(\widehat\rhob)\times \widehat\Ib_m=\widehat\qb_m$ for all $m$:
\begin{subequations}
\begin{align}
  \label{solutionq}
   \widehat{\rhob}  &\,= \,\text{diag}(\Ib_0)^{-1}
  \, \overline{\qb} &\\
  \forall m \qquad \widehat{\Ib}_{m} &\,=\,
  \text{diag}(\widehat{\rhob})^{-1}\, \widehat{\qb}_{m} &
  \label{solutionI}
\end{align}
\label{solution}
\end{subequations}
where 
%$\widehat{\Sb} \pardef{\textstyle\sum_m} \, \widehat{\qb}_{m} \,=\,
%  {\textstyle \sum_m}  \diag(\widehat{\rhob}) \, \widehat{\Ib}_{m} \,=\, M \,\diag(\widehat{\rhob})\Ib_{0}$.
\begin{align*}
\nonumber
%  \label{Contrainte}
%\begin{array}{rcl}
  \overline{\qb} &\pardef
  {\textstyle\frac{1}{M} 
    \sum_m} \, \widehat{\qb}_{m} \,=\,
 \diag(\widehat{\rhob}) \,  {\textstyle\frac{1}{M}
    \sum_m} \widehat{\Ib}_{m} 
  \,= \,\diag(\widehat{\rhob})\, \Ib_{0}.
%  \end{array}
\end{align*}
%
%leading to a solution for \eqref{critere1}
Moreover, the following implications hold:  
\begin{align*}
  \nonumber
%  \begin{array}{l}    
  I_{0;n} >0, \quad \text{and} \quad \widehat{q}_{m;n}>0, \\[0em]
  \Longrightarrow \qquad \widehat{I}_{m,n}\geq 0 \quad \text{and} 
  \quad \widehat{\rho}_n \geq 0 \qquad\forall n,m.
%  \end{array}
\end{align*}
Because we are dealing with intensity patterns, the condition
$\Ib_0\geq 0$ is always met, hence  ensuring the positivity of 
both the density and the illumination estimates. We also note 
that a solution defined by \eqref{solution} exists as long as $I_{0;n} \neq 0$ 
and  $\widehat{\rho}_n \neq 0$, $\forall n$.  The first condition
is met if the sample is illuminated everywhere, which is an obvious minimal
requirement. For any pixel sample such that $\widehat{\rho}_{n} = 0$, 
the corresponding illumination $\widehat{I}_{m;n}$ is not defined;
this is not a problem as long as the fluorescence density $\rhob$
is the only quantity of interest.
%\section{Super-resolution in Blind-SIM}
%\label{SR}

\subsection{Toward a penalized joint blind-SIM strategy}

Whereas the mechanism that conveys super-resolution with \textit{known}
structured illuminations is well understood (see~\cite{Gustafsson00}   for instance), the 
super-resolution capacity of joint blind-SIM has not been characterized yet.
It can be made clear, however, that the positivity constraint \eqref{critere1c} 
plays a central role in this regard. 
Let $\Hb^+$ be the \textit{pseudo-inverse} of  $\Hb$  
\cite[Sec. 5.5.4]{Golub96}. Then, any solution to the problem 
\eqref{critere1a}-\eqref{critere1b} [\textit{i.e.}, without positivity 
constraints] reads
\vspace{.5em}
\begin{subequations}
  \label{solutionFourier}
    \begin{align}
      \label{solutionFouriera}
      \widehat{\rhob} &=\text{diag}(\Ib_0)^{-1} (\Hb^+ \overline{\yb} + \overline{\qb}^\perp ) \\[0em]
      \label{solutionFourierb}
      \widehat{\Ib}_m &=\text{diag}(\widehat{\rhob})^{-1} (\Hb^+ \yb_m + \qb_m^\perp),  
    \end{align}
\end{subequations}
with $%
%
%\begin{equation}
%\nonumber
{\textstyle
\overline{\yb} = \frac{1}{M} \sum_m \yb_m,\quad  \text{and} \quad \overline{\qb}^\perp = \frac{1}{M} \sum_m }\qb_m^\perp
%\vspace{.25em}
%\end{equation}
%
$
where $\qb_m^\perp$ is an arbitrary element of the kernel of $\Hb$,
\ie with arbitrary frequency components above the OTF cutoff frequency.
Hence, the formulation \eqref{critere1a}-\eqref{critere1b} has no capacity to discriminate the correct high frequency components, which means that it has no super-resolution capacity.
Under the positivity constraint \eqref{critere1c}, we thus expect that
the super-resolution capacity of joint blind-SIM depends on the fact that
each illumination pattern $\Ib_m$ activates the positivity  constraint
on $\qb_m$ in a frequent manner.
Such adequate illumination patterns can be easily generated as speckle images, as proposed by~\cite{Mudry12}. 
In contrast, standard SIM rests upon the amplitude modulation of the object, \ie,   
it only needs broad-band spectra illumination patterns.

Let us stress that each problem~\eqref{critere2} is convex quadratic,
and thus admits only global solutions, which in turn provide global 
solutions to problem~\eqref{critere1}, when recombined according 
to~\eqref{solutionq}-\eqref{solutionI}. 
On the other hand, problems~\eqref{critere2} may 
not admit unique solutions, since $\left\|\yb_m -  \Hb\qb_m\right\|^2$
is not strictly convex in $\qb_m$. A simple way to enforce
unicity is to slightly modify~\eqref{critere2} by adding a strictly
convex penalization term. We are thus led to solving
\begin{equation}
  \label{critere3}
%   \begin{align}
%   \label{critere3a}
%   {\textstyle
%     \min_{\qb_m\geq0}} \,& {\textstyle \sum_{m=1}^{M} \Jc_m(\qb_m)}
% %
%   \intertext{with}
%   \label{critere3b}
%   \Jc_m(\qb) &\pardef   \left\| \yb_m - \Hb\qb\right\|^2 +   \varphi( \qb).
%   \end{align}
  {\textstyle
    \min_{\qb_m\geq0}} \, {\textstyle \sum_{m=1}^{M} \left\| \yb_m - \Hb\qb_m\right\|^2 +   \varphi( \qb_m).}
\end{equation}
Another advantage of such an approach is that $\varphi$ can be chosen
so that robustness to the noise is granted and/or some expected
features in the solution are enforced. In particular, the analysis
conveyed above suggests that favoring sparsity in each $\qb_m$ 
is suited since speckle or periodic illumination patterns tend to 
frequently cancel or nearly cancel the  components $q_{m;n}$ within
the product image $\qb_m$. 
For such illuminations,  the \textit{Near-Black Object} introduced 
in Donoho's seminal paper \cite{Donoho92a} is an appropriate 
modeling and, following this line, we found that the standard 
separable $\ell_1$ penalty provides super-resolved reconstructions 
\begin{equation}
  \label{hyperbolic}
  \varphi( \qb_m; \alpha,\beta) \pardef     \beta
   ||\qb_m||^2 + \alpha {\textstyle \sum_n |q_{m;n}|,} 
\end{equation}
with $\alpha\geq 0$ and $\beta>0$ some hyper-parameters.
With  properly tuned  $(\alpha,\beta)$, our 
penalized joint Blind-SIM strategy is expected to bring super-resolution if 
``adequate'' illumination patterns are used, \textit{i.e.,} 
they need to locally cancel out the imaged sample and their average 
$\Ib_0$ has to be known. 
\begin{figure}[t]
\centering
% \begin{tabular}{c@{\kern.4cm}c}
\begin{tabular}{c}
  \includegraphics[clip=true,bb=65 36 519 400,width=8cm]{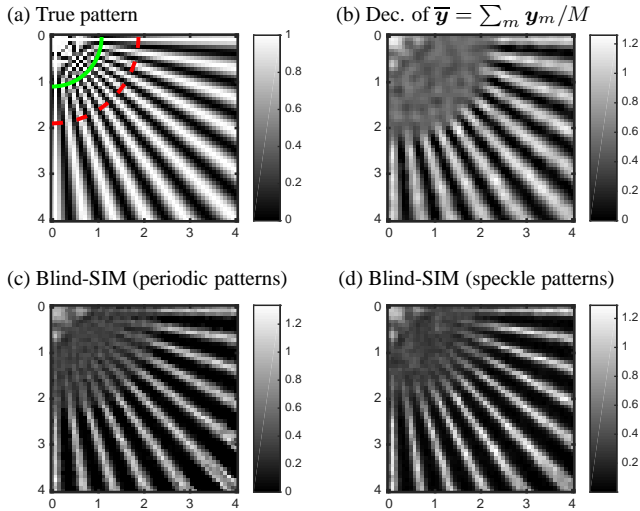}
  \psset{framesep=1pt}
  \rput[l](-4.15,3.08){\psframebox[fillstyle=solid,fillcolor=white,linecolor=white]{\footnotesize
      (d) Blind-SIM (speckle patterns)}}
 \psset{framesep=1pt}
  \rput[l](-8.56,3.08){\psframebox[fillstyle=solid,fillcolor=white,linecolor=white]{\footnotesize
      (c) Blind-SIM (periodic patterns)}}
  \psset{framesep=0pt}
  \rput[l](-4.15,6.6){\psframebox[fillstyle=solid,fillcolor=white,linecolor=white]{\footnotesize
      (b) Dec. of $\overline{\yb}=\sum_{m} \yb_m/M$}}
  \psset{framesep=1pt}
  \rput[l](-8.56,6.6){\psframebox[fillstyle=solid,fillcolor=white,linecolor=white]{\footnotesize  (a) True pattern~~~}}
  \psarc[linestyle=dashed,dash= 4pt 5pt, linecolor=red,linewidth=1.5pt](-7.9,6.32){1.15}{-90}{0}
  \psarc[linecolor=green,linewidth=1.5pt](-7.9,6.32){.66}{-90}{0}
  \end{tabular}
 %    &   
 %
%\end{tabular}
  \caption{Right lower quadrant of the ($100 \times 
    100$ pixels) true fluorescence pattern (a), positive deconvolution 
     of the averaged speckle patterns (b), and penalized joint blind-SIM
    reconstructions with $M=9$ periodic (c) and $M=200$ speckle patterns (d). 
    The graduations are in $\lambda$. The dashed (resp. solid) lines in (a)
    corresponds to the spatial frequencies transmitted by the OTF 
    support (resp. twice the OTF support).}
 \label{fig1}
\end{figure}

\subsection{Numerical illustration}
For illustrative purposes, the numerical blind-SIM experiment 
presented in~\cite{Mudry12} is now considered. The ground-truth 
consists in the 2D 'star-like' fluorescence pattern depicted in 
Fig.~\ref{fig1}(a). The $M$ collected images are simulated 
following~\eqref{eq:observation} with the PSF $h$  given by the 
Airy pattern that reads in polar $(r,\theta)$ coordinates
\begin{equation}
\label{psf}
h(r, \theta) =\left( \frac{J_1(r\,    k_0\,\text{NA})}{k_0\,r}\right)^2\,\frac{k_0^2}{\pi} 
\end{equation}
where $J_1$ is the first order Bessel function of the first kind, NA
is the objective numerical aperture set to 1.49, and $k_0 =
2\pi/\lambda$ is the free-space wavenumber with $\lambda$ 
the fluorescence emission and the excitation wavelength. The image sampling 
step for all the simulations is set to $\lambda$/12. The illumination 
set $\{\Ib_m\}_{m=1}^M$ consists either in $M=9$ \textit{periodic patterns} 
with spatial frequency equal to $2.4/\lambda$, or in $M=200$ \textit{speckle patterns} with 
spatial correlation given by \eqref{psf}. Finally, the collected
images are corrupted with Gaussian noise. The standard deviation
for a single acquisition was chosen so that the total SNR is 40 dB 
for both the periodic and speckle experiments. The subproblem 
hyperparameters were empirically set to
$(\alpha=0.3,\beta=10^{-4})$ for both the periodic  and the speckle 
illuminations. 
%The  initial-guess is $\qb_m^{(0)}=\omegab_m^{(0)}=\textbf{0}$ and the parameters
% $(\rho,\tau,\sigma)$ were set according to the tuning rule given 
%above.
%
The reconstructions of Fig.~\ref{fig1}(c)-(d) clearly show a
super-resolution effect similar to the one obtained in~\cite{Mudry12}.  
In particular, this simulation corroborates the empirical statement
that $M\approx 10$ harmonic illuminations and $M\approx 200$ 
speckle illuminations produce almost equivalent super-resolved 
reconstructions. Obviously, imaging with random speckle patterns 
remains an attractive strategy since it is achieved with a very simple 
experimental setup, see \cite{Mudry12} for details. The blind-SIM 
reconstructions shown in Fig.~\ref{fig1} were produced with 
1000 iterations of the algorithm proposed in Sec.~\ref{algo}; the total computation time for the processing 
of the ($M=200$) speckle and ($M=9$) harmonic patterns is 360 and 
17 seconds with a standard Matlab implementation on a regular computer.
On the other hand, let us remark that our strategy requires an explicit tuning of the
parameters $\alpha$ and $\beta$, whereas the 
constrained conjugate gradient approach proposed in~\cite{Mudry12} 
is regularized through the number of iterates. Since adequate values
of  $\alpha$ and $\beta$ will depend mostly on experimental parameters (PSF, 
noise and signal levels, number of views), a simple calibration step seems possible.

\section{A new optimization strategy}
\label{algo}

\begin{figure}[t]
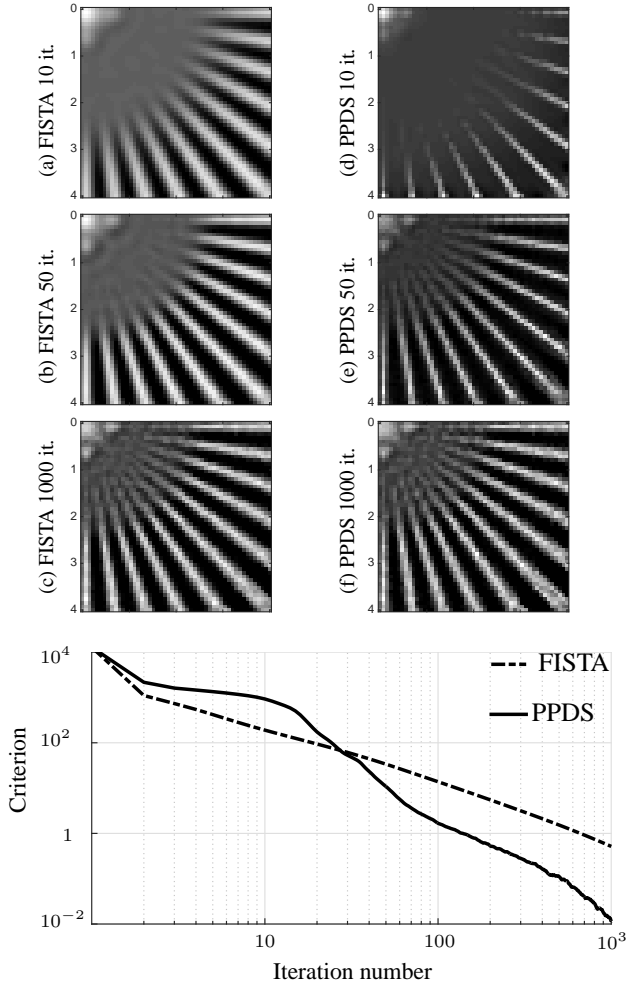

  \centering
    \begin{tabular}{c@{\kern1.25cm} c}
      \includegraphics[clip=true,bb=75 45 450 410, width=2.7cm]{FISTA_10}
      &
      \includegraphics[clip=true,bb=75 45 450 410, width=2.7cm]{CONDAT_10}\\%[.15em]
      \includegraphics[clip=true,bb=75 45 450 410, width=2.7cm]{fig1/FISTA_50}
       &
      \includegraphics[clip=true,bb=75 45 450 410, width=2.7cm]{CONDAT_50}\\%[.15em]
      \includegraphics[clip=true,bb=75 45 450 410, width=2.7cm]{FISTA_1000}
       &
      \includegraphics[clip=true,bb=75 45 450 410, width=2.7cm]{CONDAT_1000}  \end{tabular}
     \rput[l]{90}(-3.25,-3.8){\psframebox[fillstyle=solid,fillcolor=white,linecolor=white]{\footnotesize       (f) PPDS 1000 it.}}
     \rput[l]{90}(-7.25,-3.8){\psframebox[fillstyle=solid,fillcolor=white,linecolor=white]{\footnotesize       (c) FISTA 1000 it.}}
     \rput[l]{90}(-3.25,-1){\psframebox[fillstyle=solid,fillcolor=white,linecolor=white]{\footnotesize       (e) PPDS 50 it.}}
     \rput[l]{90}(-7.25,-1.){\psframebox[fillstyle=solid,fillcolor=white,linecolor=white]{\footnotesize       (b) FISTA 50 it.}}
     \rput[l]{90}(-3.25,1.75){\psframebox[fillstyle=solid,fillcolor=white,linecolor=white]{\footnotesize   (d) PPDS 10 it.}}
     \rput[l]{90}(-7.25,1.75){\psframebox[fillstyle=solid,fillcolor=white,linecolor=white]{\footnotesize       (a) FISTA 10 it.}}
    \\[1em]
%      \psfrag{15}[r]{\footnotesize  15}
%      \psfrag{100}[t]{\footnotesize  100}
%      \psfrag{1000}[t]{\footnotesize  1000}
 %     \psfrag{Proj. Grad.}{\footnotesize Projected-Gradient}
%
    ~
    \begin{tabular}{@{\kern.8cm}c}
      \psfrag{FISTA}{ FISTA}
      \psfrag{PPDS}{PPDS}
      %\psfrag{Criterion}[b]{\footnotesize Criterion}
      %\psfrag{Iteration}[br]{\footnotesize iteration number}
       \includegraphics[clip=true, bb= 70 100 522 335, width=7cm]{critere.eps} 
      %\includegraphics[clip=true, bb= 285 35 1030 340,      width=8.5cm]{convergence_fista_VS_GP_1000it_6e4_150_new.eps} 
      %\rput[l](-4.3,2.65){\psframebox[fillstyle=solid,fillcolor=white,linecolor=white]{\tiny  FISTA}}
      %\rput[l](-4.3,2.92){\psframebox[fillstyle=solid,fillcolor=white,linecolor=white]{\tiny      PPDS}}
      \rput[l](-7.7,0.1){\scriptsize      $10^{-2}$}
      \rput[l](-7.4,1.2){\scriptsize      $1$}
      \rput[l](-7.7,2.5){\scriptsize      $10^2$}
      \rput[l](-7.7,3.6){\scriptsize       $10^{4}$}
      \rput[l](-4.85,-.2){\scriptsize       $10$}
      %\rput[l](-3.25,-.2){\scriptsize       $50$}
      \rput[l](-4.6,-.6){\small  Iteration number}
      \rput[l](-2.6,-.2){\scriptsize       $100$}
      \rput[l](-.3,-.2){\scriptsize       $10^3$}
      \rput[l]{90}(-8,1.5){\small       Criterion}
      %\rput[l](-.55,-.2){\scriptsize       $10^3$}
      %\rput[l](-9.2,.1){\scriptsize       $10^8$}

      \end{tabular}\\[1.5em]
  \caption{[Top] Harmonic joint blind-SIM reconstruction of the fluorescence
    pattern achieved by the minimization of the criterion 
    \eqref{critere3} with 10, 50 or 1000 FISTA (abc)   or PPDS (def) 
    iterations. [Bottom] Criterion value (log-log scale) as a  function of the PPDS 
    (plain) or FISTA (dot) iteration number. All these
   simulations were performed with  $(\alpha=0.3,\beta=10^{-4})$. 
   %The reconstruction (f) corresponds to the one that is shown in Fig.~\ref{fig1}-(c).  
 }
  \label{fig2}
\end{figure}

We now consider the algorithmic issues involved in the constrained 
optimization problem \eqref{critere3}-\eqref{hyperbolic}. For the sake of simplicity, the 
subscript $m$  in $\yb_m$ and $\qb_m$
are dropped, however, the reader should keep in mind 
that the algorithm presented in the next section only aims at 
solving one of the $M$ sub-problems involved in the 
final joint Blind-SIM reconstruction.
At first, let us note that 
our minimization problem \eqref{critere3}-\eqref{hyperbolic} is an instance of the more general statement 
\begin{equation}
  \label{critereProx}
  {\textstyle \min_{\qb\in\eR^N} }\left[ f(\qb) \pardef g(\qb) + h(\qb) \right] 
\end{equation}
where $g$  and $h$ are closed-convex functions that may not share 
the same regularity assumptions: $g$ is supposed to be a smooth 
function with a $L$-Lipschitz continuous gradient $\boldsymbol{\nabla}
g$, but $h$ needs  not to be smooth.
Our penalized joint Blind-SIM problem hence takes the form of \eqref{critereProx} with  
\begin{subequations}
  \label{critereGen}
  \begin{align}
    \label{critereGen1}
    g(\qb) & \,=\,  || \yb - \Hb \qb||^2 + \beta ||\qb||^2 \\
%    \intertext{and}
    \label{critereGen2}  
    h(\qb) & \, = \, {\textstyle \alpha \sum_n}  \phi(q_n)%\\[.5em] % \,+\, \sum_n  I_+(q_n)    
    \intertext{where $\phi:\eR \rightarrow \eR\cup \{+\infty\}$ is such that}
   \label{critereGen3}  
    \phi (u) & \pardef \left\{
    \begin{tabular}{@{\kern1pt}ll}
      $u$ & \quad if $u\geq 0$. \\[-.2em]
      $+\infty$ & \quad elsewhere.
    \end{tabular}
  \right.
  \end{align}
\end{subequations}
Constrained non-smooth optimization problems \eqref{critereProx} 
can be solved by a dedicated proximal iteration. For that purpose, FISTA~\cite{Beck09} is recognized 
as a fast and numerically efficient alternative: let $\qb^{(0)}$ be a feasible initial-guess and 
$\omegab^{(0)}=\qb^{(0)}$, the FISTA update reads
 \begin{subequations}
  \label{algoFISTA}
\begin{align}
  \label{algo2}
&  \qb^{(k+1)} ~\longleftarrow~ \prox{}\left( \omegab^{(k)} -  \gamma  \nabla g(\omegab^{(k)})\right)\\
%\intertext{where $\gamma$ is a constant step-size and}
\label{algo3}  
&  \omegab^{(k+1)} ~\longleftarrow~\qb^{(k+1)} + {\textstyle \frac{k-1}{k+2}}  \big(\qb^{(k+1)} - \qb^{(k)}\big)
  \end{align}
\end{subequations}
with $\gamma>0$ and $\prox{}$ the proximity operator 
(or \textit{Moreau envelope}) of the function $\gamma h$  
that is easily found from (\ref{critereGen2}-c)  
\begin{equation}  
  \label{eq:prox2}
  \prox{}  (\omegab) =  \textbf{vect}\left(\max\{\omega_n - \gamma \alpha,0\}\right). 
\end{equation}
Global convergence of \eqref{algoFISTA} is granted
provided that the step-size is such that $0 <\gamma < 1/L$ 
with 
%
%\begin{equation}
%\label{Lipschitz}
$L = 2 \left(\lambda_{\text{max}}(\Hb^t \Hb) + \beta\right)$
%\end{equation}
%
where $\lambda_{\text{max}}(\Ab)$ denotes the highest eigenvalue of
the matrix $\Ab$.  
The decreasing rate achieved by $f(\qb^{(k)})$ is $O(1/k^2)$,
which is a substantial gain compared to the $O(1/k)$ rate
of the standard proximal iteration \cite{Beck09}. For our specific
problem, however, the convergence speed of FISTA is too 
low. Figure~\ref{fig2}  gives an illustration of this issue: FISTA 
requires almost one thousand iterations to converge. This is
precisely the motivation for the introduction of the \textit{preconditioned 
primal-dual splitting} (PPDS), a new algorithm that rests on the
very versatile splitting technique proposed in \cite{Condat13}.
Due to space limits, however, the comprehensive presentation 
of this iterative strategy is not detailed. 
Let $(\tau,\rho,\sigma)$ be a triplet of positive values, the PPDS iteration 
consists in a primal-dual update that reads 
\begin{subequations}
  \label{iterationfinal}
    \begin{align}
    \label{iterationfinala}
    &   \qb^{(k+1)} \longleftarrow \qb^{(k)} - \rho \tau  \Bb \zetab^{(k)} \\ 
    \label{iterationfinalb}  
    &  \omegab^{(k+1)} \longleftarrow \omegab^{(k)} +\rho [
     % \prox{\sigma h^\star} ( \deltab^{(k)})  - \omegab^{(k)}\right]
        \prox{\star} ( \deltab^{(k)})  - \omegab^{(k)}]
    \end{align}
\end{subequations}
with $\zetab^{(k)}:= \boldsymbol{\nabla} g(\qb^{(k)}) + \omegab^{(k)}$
and where the \textit{preconditioning matrix} $\Bb$ is chosen from the Geman and 
Yang \textit{semi-quadratic} construction \cite{Geman95}, \cite[Eq. (6)]{Allain06b} 
\begin{equation}
  \label{Precondbis}
  \Bb := \left( 2\Hb^t \Hb + 2\beta\,\Ib/a\right)^{-1}
\end{equation}
with $a>0$ a free parameter. In the dual update
\eqref{iterationfinalb},  we noted $\deltab^{(k)} :=\omegab^{(k)} + \sigma ( \qb^{(k)} - 2 \tau
\Bb\zetab^{(k)})$ and
% $\prox{\sigma h^\star} (\omegab) = \omegab -\sigma \prox{h/\sigma} (\omegab/\sigma)$ the proximal mapping
%applied to  the \textit{Fenchel transform} of $\sigma h$ that reads from (\ref{critereGen2}-c)
%
\begin{equation}  
  \label{eq:prox3}
%  \prox{\sigma h^\star}  (\omegab) =  \textbf{vect} \left( \text{min}
  \prox{\star}  (\omegab) =  \textbf{vect} \left( \text{min}
    \left\{\omega_n , \alpha\right\} \right).
\end{equation}
The convergence of the sequence defined by \eqref{iterationfinal}  is granted if
some conditions are met for the parameters $(\rho,\tau,\sigma,a)$. Within the 
convergence domain ensured by \cite[Theorem 5.1]{Condat13}, the practical 
tuning  of these parameters is somewhat tricky as it may dramatically impair the 
convergence speed. An \textit{automatic} and efficient tuning strategy
ensuring convergence has been devised;  the PPDS reconstructions 
shown in Figure~\ref{fig2}  were obtained with this tuning. 
%
% The following tuning rule has been found efficient
% in practice: let $\Lc$ be the Lipschitz constant\footnote{For sake of
%   completness, we note that the expression of $\Lc$ depends on the
%   value of the parameter $a$. Its numerical value, however, is easily 
% obtained in any case, see \cite{TechRep16} for details.} of $\boldsymbol{\nabla}
% \Gc$ with $\Gc(\vb) := g(\Bb^{1/2}\vb)$, we set $\sigma=
% \lambda^{-1}_{\text{max}}(\Bb)\times \Lc/2$, $\tau=1/\Lc$ and $\rho=0.99$. 
%
% Finally, the parameter $a$ in \eqref{Precondbis} needs also  to be
% set. Clearly, the preconditioning matrix $\Bb=(\boldsymbol{\nabla}^2 g)^{-1}$
% is obtained with $a=1$, hence correcting the curvature anisotropies induced 
% by  $g$ in the primal update \eqref{iterationfinala}. In practice, a
% faster convergence speed was obtained with $a\approx 10\times \beta$; 
% the PPDS reconstructions shown in Figure~\ref{fig2}  were obtained with this tuning.
%
The interest of the proposed strategy is that the iteration
\eqref{iterationfinal} can be easily implemented at a low computational
cost. Indeed, since $\boldsymbol{\nabla} g$ is a linear function, the
primal step \eqref{iterationfinala} can be transposed 
in the Fourier domain, and it is easy to show that the 
computational burden of the PPDS iteration is dominated by 
only one FFT/iFFT pair. In other words, for our problem
\eqref{critereGen}, FISTA and PPDS share the \textit{same} complexity 
per iteration.

\section{Conclusion}
The reformulation presented in Sec.~\ref{Reformulation}  unveiled some
of the super resolution properties of the joint reconstruction problem  
introduced in \cite{Mudry12}. We feel however that this joint blind-SIM
approach deserves further investigations, both from the theoretical 
and the experimental viewpoints.
%  In particular, the \textit{joint} 
% reconstruction approach considered here produces some systematic 
% errors  (\textit{i.e.}, bias) that should be evaluated. Indeed, we are currently exploring a \textit{marginal
% strategy} aiming at estimating $\rhob$ \textit{only}, which could be preferable from the statistical viewpoint.
%
%Some experimental datasets should be considered shortly.  
In particular, one expected difficulty arising in the processing of 
real data sets is the strong background induced in the focal plane 
by the out-of-focus light. This phenomenon prevents the local 
extinction of the excitation
intensity, hence destroying the expected super-resolution in joint blind-SIM. 
The modeling of this background with a very smooth function is possible
\cite{Orieux12} and will be considered. A different approach would be to
solve the reconstruction problem in its 3D structure, which is
numerically challenging, but remains a mandatory step to achieve 
3D reconstructions.     
Finally, the PPDS strategy introduced in this communication is a 
promising optimization tool that will be tested with
other applications. In particular, the deconvolution under a 
positivity constraint  for the standard SIM problem should be
considered shortly.% \cite{Boulanger14}. 
\medskip

\noindent%
\textbf{Acknowledgement.} The GDR ISIS and the Agence Nationale de la
Recherche (ANR-12-BS03-0006) are  acknowledged for the partial funding of this work.   

\bibliographystyle{plain}
\bibliography{fiche}

% \begin{thebibliography}{99}

% \bibitem{companion}
% M.~Goossens, F~Mittelbach et A.~Samarin.
% \emph{The \LaTeX{} Companion}.
% Addison-Wesley, 1994.

% \bibitem{lamport94a}
% L.~Lamport.
% \emph{\LaTeX{} User's Guide and Reference Manual}.
% Addison-Wesley, 1994.

% \end{thebibliography}

\end{document}